\documentclass[conference]{IEEEtran}
\IEEEoverridecommandlockouts
\newtheorem{Theorem}{Theorem}
\newtheorem{Lemma}{Lemma}

\newtheorem{defi}{Definition}

\usepackage{algorithm}
\usepackage{algorithmic}
\usepackage{graphicx}
\usepackage{textcomp}
\usepackage{stfloats}
\usepackage{subfig}
\usepackage{amssymb,amsmath,bm}
\usepackage{cite}
\usepackage{times}
\usepackage{latexsym}
\usepackage{array}
\usepackage{fancyhdr}
\usepackage{psfrag}
\usepackage{multirow}
\usepackage{xcolor}
\usepackage{color}
\usepackage{makecell}
\usepackage{threeparttable}
\usepackage{hyperref}

\begin{document}
\title{Energy~Efficiency~of~MIMO~Massive~Unsourced Random Access with Finite Blocklength\\}

\author{\IEEEauthorblockN{Junyuan~Gao,
Yongpeng~Wu,~\IEEEmembership{Senior~Member,~IEEE,}
Tianya~Li,
and~Wenjun~Zhang,~\IEEEmembership{Fellow,~IEEE}}
\thanks{The work of Y. Wu was supported in part by the National Key R\&D Program of China under Grant 2018YFB1801102, National Science Foundation of China (NSFC) under Grant 62122052 and 62071289, 111 project BP0719010, and STCSM 18DZ2270700.
\emph{(Corresponding author: Yongpeng~Wu)}}
\thanks{J. Gao, Y. Wu, T. Li, and W. Zhang are with the Department of Electronic Engineering, Shanghai Jiao Tong University, Minhang 200240, China (e-mail: \{sunflower0515, yongpeng.wu, tianya, zhangwenjun\}@sjtu.edu.cn)}
}

\maketitle

\begin{abstract}
  This paper investigates the energy efficiency of massive unsourced random access~(URA) in multiple-input multiple-output quasi-static Rayleigh fading channels.
  Specifically, we derive achievability and converse bounds on the minimum required energy-per-bit under the per-user probability of error constraint,
  where the converse bounds contain two parts: one is general and the other is a weaker ensemble bound.
  Numerical evaluation shows that the gap between our achievability and converse bounds is less than $5$~dB in the considered regime.
  Some practical schemes are energy-inefficient compared with our bounds especially when there are many users.
  Moreover, we observe that in contrast to the sourced random access paradigm, the URA paradigm achieves higher spectral efficiency.
\end{abstract}

\section{Introduction} \label{section1}

  As a typical use case in future wireless networks,
  massive machine-type communication has two distinct features different from traditional human-type communication~\cite{wuyp}.
  First, there are a large number of users, while only a fraction of them are active at any given time.
  Second, active users transmit small data payloads to the base station~(BS) with stringent latency and energy constraints.
  Massive random access technology has been proposed for this scenario, which includes sourced and unsourced random access~(SRA and URA) paradigms. 
  For SRA, the BS requires to identify active users and decode their messages.
  For URA, the BS is only interested in the transmitted messages but not users' identities.
  The URA paradigm was introduced in~\cite{A_perspective_on} and has attracted great attention, calling for new information-theoretic analysis.

  On this topic, finite-blocklength~(FBL) bounds on the minimum required energy-per-bit were derived in~\cite{A_perspective_on} and~\cite{RAC_fading} for URA in Gaussian and Rayleigh fading channels, respectively, under the per-user probability of error~(PUPE) constraint and the assumption of knowing the number $K_a$ of active users.
  Further, in~\cite{noKa}, the result in~\cite{A_perspective_on} was extended to the setting with unknown $K_a$.
  Notably, the above-mentioned FBL results are established for the setting with a single BS antenna.
  Indeed, the use of large antenna arrays has great benefits.
  Specifically, it was proved in~\cite{Caire1} that with $n$ channel uses and $L$ BS antennas satisfying $K_a/L=o(1)$, up to $K_a \!=\! \mathcal{O}(n^2)$ active users can be identified from $K\!=\!\Theta(K_a)$ potential users, but it reduces to $K_a \!=\! \mathcal{O}(n)$ when there is a single BS antenna.

  In this paper, we investigate the energy efficiency of URA in multiple-input multiple-output~(MIMO) quasi-static Rayleigh fading channels.
  Assuming all users share a common codebook, 
  we derive achievability and converse bounds on the minimum required energy-per-bit under a PUPE constraint.
  Specifically, we utilize random coding and maximum likelihood~(ML) decoding to derive the achievability bound, where Fano's ``good region'' technique~\cite{1961} is applied since the error event is the union of many events.
  Our converse bounds~contain two parts, namely the single-user bound and multi-user Fano-type bound. The former is general and the latter is limited to Gaussian codebooks.
  Numerical results verify the tightness of our bounds and indicate their importance to benchmark practical schemes.
  Some schemes are shown to be suboptimal especially in large $K_a$ regime.
  Moreover, in contrast to SRA, the URA paradigm achieves higher spectral efficiency.

  \emph{Notation:} Throughout this paper, uppercase and lowercase boldface letters denote matrices and vectors, respectively.
  We use $\left[\mathbf{x} \right]_{m}$ to denote the $m$-th element of $\mathbf{x}$ and $\left[\mathbf{X} \right]_{m,n}$ to denote the $\left( m,n \right)$-th element of $\mathbf{X}$.
  We use $\left(\cdot \right)^{T}$, $\left(\cdot \right)^{H}$, $\left|\mathbf{X}\right|$, $\left\|\mathbf{x} \right\|_{p}$, and $\left\|\mathbf{X} \right\|_{F}$ to denote transpose,
  conjugate transpose, determinant, ${\ell}_p$-norm, and Frobenius norm, respectively.
  Let $\operatorname{diag} \left\{ \mathbf{x} \right\}$ denote a diagonal matrix with $\mathbf{x}$
  comprising its diagonal elements and $\operatorname{diag} \left\{ \mathbf{A}, \mathbf{B} \right\}$ denote a block diagonal matrix.
  We use $\cdot\backslash\cdot$ to denote set subtraction and $\left| \mathcal{A} \right|$ to denote the cardinality of a set $\mathcal{A}$.
  For an integer $k\!>\!0$, we denote $[k]\!=\!\left\{1,\ldots,k \right\}$.
  We use $\mathcal{CN}(\cdot ,\cdot)$ and $\chi^2(\cdot)$ to denote the circularly symmetric complex Gaussian distribution and central chi-square distribution, respectively.
  We use $\gamma\left(\cdot, \cdot\right)$ and $\Gamma\left(\cdot\right)$ to denote the lower incomplete gamma function and gamma function, respectively.
  The complement of event $\mathcal{G}$ is denoted as $\mathcal{G}^c$.
  For $0 \leq p \leq 1$, we denote~$h_2(p) = -p\log_2 p -(1-p)\log_2(1-p)$. 


\section{System Model} \label{section2}
  We consider an uplink single-cell system consisting of a BS equipped with $L$ antennas and $K$ single-antenna users, where only $K_a$ users are active due to sporadic traffic.
  The active user set is denoted as $\mathcal{K}_a$.
  Each active user has a message of $J = \log_2 M$ bits to transmit with $n$ channel uses.

  We consider a quasi-static Rayleigh fading channel model.
  The $l$-th antenna of the BS observes $\mathbf{y}_l$ given by
  \begin{equation} \label{eq_yl}
    \mathbf{y}_l = {\sum}_{k\in{\mathcal{K}}_a}{h}_{k,l} \mathbf{x}_{W_k} +\mathbf{z}_l \in \mathbb{C}^{n},
  \end{equation}
  where ${h}_{k,l} {\stackrel{\text{i.i.d.}}{\sim}} \mathcal{CN}(0,1)$ denotes the fading coefficient between the $k$-th user and the $l$-th BS antenna;
  the noise vector is $\mathbf{z}_l\!\sim\!\mathcal{CN}(\mathbf{0}, \mathbf{I}_n)$;
  $W_{k}$ denotes the message of active user~$k$, which is chosen uniformly at random from $[M]$;
  the transmitted codeword of active user~$k$ is denoted as $\mathbf{x}_{W_k}$.
  Here, we assume all users share a common codebook, and the matrix $\mathbf{X} = \left[\mathbf{x}_{1},\ldots, \mathbf{x}_{M}\right] \in \mathbb{C}^{n\times M}$ is obtained by concatenating all codewords.
  The received signal $\mathbf{Y} = \left[ \mathbf{y}_1, \ldots, \mathbf{y}_L \right] $ is given~by
  \begin{equation} \label{eq_y}
    \mathbf{Y} =\mathbf{X}\boldsymbol{\Phi}\mathbf{H}+\mathbf{Z} \in \mathbb{C}^{n\times L}  ,
  \end{equation}
  where the binary selection matrix $\boldsymbol{\Phi}\!\in\!\{0,1\}^{M \times K}$ satisfies that $\left[\boldsymbol{\Phi}\right]_{W_k,k} = 1$ if the $k$-th user is active and transmits the $W_{k}$-th codeword, and $\left[\boldsymbol{\Phi}\right]_{W_k,k} = 0$ otherwise;
  $\mathbf{H} = \left[\mathbf{h}_1, \ldots, \mathbf{h}_L \right] \in \mathbb{C}^{K\times L}$ with $\mathbf{h}_l = \left[ {h}_{1,l}, \ldots,{h}_{K,l} \right]^T \in \mathbb{C}^{K}$ for $l\in[L]$.

  We assume neither the BS nor users know the instantaneous channel state information~(CSI) in advance, but they both know the distribution.
  As in~\cite{A_perspective_on}, $K_a$ is assumed to be known to the BS in this work due to space constraints, and our results can be extended to the scenario without known $K_a$ applying similar ideas in~\cite{noKa}.
  Next, we introduce the notion of the URA code:\begin{defi}[\cite{A_perspective_on}]\label{defi:common_codebook}
    Let $\mathcal{X}$, $\mathcal{H}_k$, and $\mathcal{Y}$ denote the input alphabet of active users, the channel fading coefficient alphabet of active user $k$, and the output alphabet for the channel~\eqref{eq_y}, respectively.
    An $(n,M,\epsilon,P)$ massive URA code consists of
    \begin{enumerate}
      \item An encoder $\emph{f}_{\text{en}}: [M] \mapsto \mathcal{X}$ that maps the message $W_k$ to a codeword $\mathbf{x}_{W_k} \in \mathcal{X}$ for $k\in\mathcal{K}_a$, where $W_k$ is chosen independently and uniformly from $[M]$ for $k\in\mathcal{K}_a$.
          The codewords in $\mathcal{X}$ satisfy the maximum power constraint
          \vspace{-0.1cm}
          \begin{equation}\label{eq:power_constraint}
            \left\|\mathbf{x}_{m}\right\|_{2}^{2} \leq nP,\;\;\;\; m\in[M].
          \end{equation}
      \item A decoder $\emph{g}_{\text{de}}: \mathcal{Y} \!\mapsto\! \binom{[M]}{K_a}$ satisfing the PUPE constraint
          \begin{align}
            P_{e} = \frac{1}{K_a} {\sum}_{k\in {\mathcal{K}_a}} & \mathbb{P}\left[ \left\{W_{k} \notin \hat{\mathcal{W}} \right\} \right. \notag \\
            & \!\!\!\!\!\!\!\!\!\!\!\! \left.\cup\left\{W_{k}=W_{i} \text { for some } i \neq k\right\} \right] \leq \epsilon .\label{PUPE}
          \end{align}
          Here, $\!\hat{\mathcal{W}}$ denotes the set of decoded messages of size~$\!K_a$.
    \end{enumerate}
  \end{defi}

  Let $S_e\!=\!\frac{K_a J}{n}$ denote the spectral efficiency and $E_b \!=\!\frac{nP}{J}$ denote the energy-per-bit. The minimum required energy-per-bit is denoted as $E^{*}_{b}(n,M,\epsilon) = \inf \left\{E_b: \exists (n,M,\epsilon,P) \text{ code} \; \right\}$.

\section{Main results} \label{section3_sub2_subsub1}
  An achievability bound on the minimum required energy-per-bit for URA in MIMO channels is given in Theorem~\ref{Theorem_noCSI_achi}.\begin{Theorem} \label{Theorem_noCSI_achi}
    The minimum required energy-per-bit for the URA model described in Section~\ref{section2} can be upper-bounded as
      \begin{equation}
        E^{*}_{b}(n,M,\epsilon) \leq \inf \frac{n {P}}{J}.
      \end{equation}
    Here, the $\inf$ is taken over all $P > 0$ satisfying that
      \begin{equation} \label{eq_noCSI_PUPE}
        \epsilon \geq \min_{0< P'< P} \left\{ p_0 + {\sum}_{t=1}^{K_a} \frac{t}{K_a} \min \left\{ 1, p_{t} \right\} \right\},
      \end{equation}
    where
    \begin{equation} \label{eq_noCSI_p0}
      p_0 = \frac{\binom{K_a}{2}}{M} + K_a \left( 1 - \frac{\gamma\left(n, \frac{nP}{P'}\right)}{\Gamma\left(n\right)}\right),
    \end{equation}
    \begin{equation} \label{eq_noCSI_pt}
      p_t = \min_{ 0 \leq \omega \leq 1, 0 \leq \nu }
       \left\{ q_{1,t}\left(\omega,\nu\right)
       + q_{2,t}\left(\omega,\nu\right) \right\}  ,
    \end{equation}
    \vspace*{-5mm}
    \begin{align}\label{eq_noCSI_q1t}
      & q_{1,t} \!\left(\omega,\nu\right) \notag\\
      & \!=\!  \binom{ K_a }{t}  \binom{M\!-\!K_a }{t} \mathbb{E}_{\mathbf{C}} \!\left[
      \min_{ {u\geq 0,r\geq 0, \lambda_{\min}\left(\mathbf{B}\right) > 0}}
      \!\! \exp \!\left\{  L rn\nu  \right\} \right. \notag \\
      &  \!\!\!\;\;\; \cdot \exp \!\left\{  L\! \left( (u\!-\!r) \!\ln \!\left|\mathbf{F}\right|
      - \!u \!\ln \!\left| {\mathbf{F}_2} \right|
      + \!r\omega \!\ln \!\left| \mathbf{F}_{1} \right|
      \!-\! \ln \!\left| \mathbf{B} \right| \right)
      \right\} \!\!\bigg] ,\!
    \end{align}
    \vspace*{-5mm}
    \begin{align}
      & q_{2,t}\!\left(\omega,\nu\right)
      \notag \\
      & =  \min_{ \delta\geq0 } \!\binom {K_a\!} {t} \mathbb{E}_{\mathbf{C}} \! \left[ \frac{\gamma\left( Lm,  c_{\delta} \right)}{\Gamma\left( Lm \right)}
      \!+\! 1 \!-\! \frac{\!\gamma\left( nL, nL\! \left( 1\!+\!\delta \right)\right)}{\Gamma\left( nL \right)} \right]\!, \label{eq_noCSI_q2t}
    \end{align}
    \begin{equation}
      c_{\delta} =  \frac{ L \left( n(1+\delta)(1-\omega) - \omega\ln\left|\mathbf{F}_1\right| + \ln\left|\mathbf{F}\right|  - n\nu \right) }
      {\omega \prod_{i=1}^{m}\lambda_i^{{1}/{m}}  },
    \end{equation}
    \begin{equation}
      \mathbf{B} = (1-u+r) \mathbf{I}_n
      + u \mathbf{F}_2^{-1} \mathbf{F}
      - r \omega \mathbf{F}_{ 1}^{-1} \mathbf{F},
    \end{equation}
    \begin{equation}
      \mathbf{F} = \mathbf{I}_n+ \mathbf{C}\boldsymbol{\Gamma}_{ S_{\mathcal{K}_a} }\mathbf{C}^H,
    \end{equation}
    \begin{equation}
      \mathbf{F}_1 = \mathbf{I}_n+ \mathbf{C}\boldsymbol{\Gamma}_{ S_{\mathcal{K}_a} \backslash {S}_1 }\mathbf{C}^H,
    \end{equation}
    \begin{equation}
      \mathbf{F}_2  = \mathbf{I}_n+ \mathbf{C} {\boldsymbol{\Gamma}}_{ S_{\mathcal{K}_a} \backslash  S_1 \cup S_2 }  \mathbf{C}^H.
    \end{equation}
  Here, $S_{\mathcal{K}_a}$ is an arbitrary $K_a$-subset of $[M]$; 
  ${S}_{1}$ is an arbitrary $t$-subset of $S_{\mathcal{K}_a}$;
  $S_{2}$ is an arbitrary $t$-subset of $[M] \backslash S_{\mathcal{K}_a}$;
  $\mathbf{C}\in\mathbb{C}^{n\times M}$ has i.i.d. $\mathcal{CN}\left(0, P'\right)$ entries;
  ${\boldsymbol{\Gamma}}_{S} = \operatorname{diag} \left\{ {\boldsymbol{\gamma}}_{S} \right\}\in \left\{ 0,1 \right\}^{M\times M}$ for 
  $S\!\subset\![M]$, where $\left[ {\boldsymbol{\gamma}}_{S} \right]_{i} \!=\! 1$ if $i\!\in\! S$ and $\left[ {\boldsymbol{\gamma}}_{S} \right]_{i} \!=\! 0$ otherwise;
  the eigenvalues of $\mathbf{F}_1^{-1} \mathbf{C}\boldsymbol{\Gamma}_{S_1}\mathbf{C}^H$ of rank $m\!=\!\min\!\left\{ n,t \right\}$ are denoted as $\lambda_1, \ldots, \lambda_n$ in decreasing order.
  \begin{IEEEproof}
  We use random coding to generate a common codebook $\mathcal{C}$, where the codeword $\mathbf{c}_{m}\!\stackrel{\mathrm{i.i.d.}}{\sim} \! \mathcal{CN}\!\left(0,P'\mathbf{I}_{n}\right)$ with $P'\!< P$ for $m \in [M]$.
  Let $\mathbf{C} \;\!\!=\;\!\! \left[ \mathbf{c}_{1}, \ldots, \mathbf{c}_{M} \right]$.
  If user $k$ is active, it transmits $\mathbf{x}_{W_k} = \mathbf{c}_{W_k} 1 \big\{ \left\|\mathbf{c}_{W_k}\right\|_{2}^{2} \leq n P \big\}$.~The transmitted messages of active users, i.e. ${S}_{\mathcal{K}_a} = \left\{W_{k}: k\in\mathcal{K}_a\right\}$, are sampled independently with replacement from $[M]$.
  To upper-bound the PUPE, we perform two changes of measure: 1)~the messages in ${S}_{\mathcal{K}_a}$ are sampled uniformly without replacement from $[M]$;
  2)~the active user~$k$ transmits $\mathbf{x}_{W_k} = \mathbf{c}_{W_k}$.
  The total variation distance between the true measure and the new one is bounded by $p_0$ in~\eqref{eq_noCSI_p0}.
  Thus, we can bound the PUPE~as
  \begin{equation} \label{eq_PUPE_upper_noCSI}
    P_{e} \leq {\sum}_{t=1}^{K_a} \frac{ t}{K_a} \mathbb{P} \left[ \mathcal{F}_t \right] + p_0 ,
  \end{equation}
  where $\mathcal{F}_t$ denotes the event that there are exactly $t$ misdecoded messages under the new measure and is bounded as follows.

  We use the ML decoder to obtain the estimated set $\hat{S}_{\mathcal{K}_a}$ of the transmitted messages.
  The decoder outputs
  \begin{equation}
    \hat{S}_{\mathcal{K}_a} = \big\{\; \emph{f}_{\text{en}}^{-1}\left( \hat{\mathbf{c}} \right): \hat{\mathbf{c}} \in \hat{\mathcal{C}}_{\mathcal{K}_a} \big\},
  \end{equation}
  \begin{equation}\label{eq:decoderoutput_noCSI}
    \hat{\mathcal{C}}_{\mathcal{K}_a}
    =\arg \min_{ \hat{\mathcal{C}}_{\mathcal{K}_a} \subset \mathcal{C}: \left| \hat{\mathcal{C}}_{\mathcal{K}_a} \right| = K_a }
    g\;\!( \hat{\boldsymbol{\Gamma}} ) ,
  \end{equation}
  where $\hat{\boldsymbol{\Gamma}} = \operatorname{diag} \left\{ \hat{\boldsymbol{\gamma} } \right\}\in\{0,1\}^{M\times M}$ with $[\hat{\boldsymbol{\gamma} }]_i=1$ if $\mathbf{c}_i \in \hat{\mathcal{C}}_{\mathcal{K}_a}$
  and the log-likelihood cost function $g\;\!( \hat{\boldsymbol{\Gamma}} )$ is given by
  \begin{equation}\label{eq:g_noCSI}
    g ( \hat{\boldsymbol{\Gamma}} )
    \!=\! L \ln \big| \mathbf{I}_n + \mathbf{C}\hat{\boldsymbol{\Gamma}}\mathbf{C}^H \big|
    +  \operatorname{tr}\!\left(  \mathbf{Y}^{\!H}
    \big( \mathbf{I}_n\!+\! \mathbf{C}\hat{\boldsymbol{\Gamma}}\mathbf{C}^H \big)^{\! -1}   \mathbf{Y}  \right)\!.\!
  \end{equation}

  Let $S_1 \!\subset\! S_{\mathcal{K}_a}$ denote the set of misdecoded messages.
  Let $S_2 \!\subset\! [M] \backslash S_{\mathcal{K}_a}$ denote the set of false-alarm messages.
  We rewrite ``$\cup_{ S_1 \subset S_{\mathcal{K}_a}, \left| S_1 \right| = t }$'' to ``$\cup_{S_1}$'' and ``$\cup_{ S_2 \subset [M] \backslash S_{\mathcal{K}_a}, \left| S_2 \right| = t }$'' to ``$\cup_{S_2}\!$'' for brevity; and similarly for $\!\sum\!$ and $\cap$.
  Then, we have
    \begin{equation}
      \mathbb{P} \left[ \mathcal{F}_t  \right]
      \!\stackrel{(a)}{\leq}\! \mathbb{P} \left[ \mathcal{G}_{e}  \right]
      \!\stackrel{(b)}{\leq}\!\!\min_{ {{0 \leq \omega \leq 1,\nu\geq0}} } \!\left\{ \mathbb{P} \!\left[
      \mathcal{G}_{e}
      \cap \mathcal{G}_{\omega,\nu} \right]
      + \mathbb{P} \!\left[ \mathcal{G}_{\omega,\nu}^c \right] \right\} \!\label{eq:noCSI_pft2_goodregion} ,
    \end{equation}
  where $\mathcal{G}_{e}  \!=\! \bigcup_{S_{1},S_{2}}
      \! \left\{ g \!\left( {\boldsymbol{\Gamma}}_{ \!S_{\mathcal{K}_a} \!\backslash S_1 \cup S_2}   \right)
      \!\leq\! g \!\left( \boldsymbol{\Gamma}_{\!S_{\mathcal{K}_a}} \!\right)  \right\}$
  and $\mathcal{G}_{\omega,\nu} \!=\! \bigcap_{S_1}  \!\left\{ \mathbf{Y} \!\in\! \mathcal{R}_{t,S_1} \right\}$.
  Here, (a) holds because we treat all ties in $\mathcal{G}_{e}$ as errors;
  (b) follows from Fano's ``good region'' technique given in Lemma~\ref{Lemma:gr}, where the ``good region'' $\mathcal{R}_{t,S_1}$ is chosen~as
  \begin{equation} \label{eq:goodregion_noCSI}
    \mathcal{R}_{t,S_1} = \left\{ \mathbf{Y} : g\left( \boldsymbol{\Gamma}_{S_{\mathcal{K}_a}} \right) \leq \omega g\left( {\boldsymbol{\Gamma}}_{S_{\mathcal{K}_a}\backslash S_1 } \right) + nL\nu \right\}.
  \end{equation}
  \vspace*{-4mm}
  \begin{Lemma}[\cite{1961}]\label{Lemma:gr}
    Let $\mathbf{y}$ be the received signal and $\mathcal{R}$ be a ``good region''.
    We can bound the decoding error probability~as
    \begin{equation} \label{eq:lemma_gr}
      \mathbb{P}\;\![e] \leq \mathbb{P}\;\![ e \cap \{\mathbf{y}\in \mathcal{R}\} ]
      + \mathbb{P}\;\![ \mathbf{y}\notin \mathcal{R} ].
    \end{equation}
  \end{Lemma}

  The term $\mathbb{P}  \left[ \mathcal{G}_{e} \cap \mathcal{G}_{\omega,\nu} \right]$ in~\eqref{eq:noCSI_pft2_goodregion} can be further bounded~as
    \begin{align}
      & \mathbb{P}  \left[ \mathcal{G}_{e} \cap \mathcal{G}_{\omega,\nu} \right] \notag \\
      & \leq {\sum}_{S_1,S_2} \mathbb{E}_{\mathbf{C}}\! \left[ \mathbb{P} \!\left[ \left\{  g \!\left( {\boldsymbol{\Gamma}}_{S_{\mathcal{K}_a} \backslash S_1 \cup S_2}   \right)
      \leq g \!\left(\boldsymbol{\Gamma}_{S_{\mathcal{K}_a}} \right)  \right\} \right. \right. \notag\\
      & \;\;\;\;\;\;\;\;\;\;\;\;\;\;\;\; \left. \left. \left.
      \cap \left\{ g \!\left(\boldsymbol{\Gamma}_{S_{\mathcal{K}_a}} \right) \leq
      \omega g \!\left(  {\boldsymbol{\Gamma}}_{ S_{\mathcal{K}_a} \backslash S_1 } \right) \!+\! nL\nu \right\}
      \right| \mathbf{C} \right] \right] \label{eq_noCSI_q1t_union} \\
      & \leq \!{\sum}_{S_1,S_2}
      \!\mathbb{E}_{\mathbf{C}} \Big[ \min_{ {0\leq u,0\leq r} } \!
      \mathbb{E}_{\mathbf{H},\mathbf{Z}} \! \left[ \exp \left\{
      rnL\nu
      + (u-r) g \!\left(\boldsymbol{\Gamma}_{S_{\mathcal{K}_a}} \right) \right. \right.  \notag \\
      & \;\;\;\;\;\;\;\;\;\;\;\;\;\;\;   \left. \left. \left.
       - u g\!\left( {\boldsymbol{\Gamma}}_{ S_{\mathcal{K}_a} \! \backslash S_1 \cup S_2}   \right)
       + r\omega g\!\left(  {\boldsymbol{\Gamma}}_{ S_{\mathcal{K}_a} \!\backslash S_1 } \right) \right\}       \right| \mathbf{C} \right] \Big] \label{eq_noCSI_q1t_chernoff}  \\
      & \leq q_{1,t} \!\left(\omega,\nu\right) , \label{eq:noCSI_q1t_exp_y}
    \end{align}
  where $q_{1,t}\!\left( \omega, \nu \right)$ is given in~\eqref{eq_noCSI_q1t};
  \eqref{eq_noCSI_q1t_union} follows from the union bound;
  \eqref{eq_noCSI_q1t_chernoff} follows by applying the Chernoff bound $\mathbb{P}\left[\{Z \geq 0\} \!\cap\! \{W \geq 0\}\right] \leq \mathbb{E}\left[\exp\left\{u Z\!+\!r W\right\}\right]$ for $u,r \geq 0$~\cite{elements_IT} to the probability in~\eqref{eq_noCSI_q1t_union};
  \eqref{eq:noCSI_q1t_exp_y} follows by applying Lemma~\ref{lemma:expectation_bound2} to the expectation in~\eqref{eq_noCSI_q1t_chernoff} over $\mathbf{H}$ and $\mathbf{Z}$.

  \begin{Lemma}[\cite{quadratic_form1}]\label{lemma:expectation_bound2}
    Assume that ${\mathbf{x}} \in \mathbb{C}^{p\times 1}$ is distributed as $ {\mathbf{x}} \sim \mathcal{CN}\left( \mathbf{0},  {\boldsymbol{\Sigma}} \right)$.
    Let $\mathbf{B}\in\mathbb{C}^{p\times p}$ be a Hermitian matrix.
    For any $\gamma$, if the eigenvalues of $\mathbf{I}_{p} - \gamma {\boldsymbol{\Sigma}}\mathbf{B}$ are positive, we have
    \begin{equation}\label{eq:lemma_expectation}
      \mathbb{E}\left[ \exp \left\{ \gamma \mathbf{x}^{H}\mathbf{B}\mathbf{x} \right\} \right] = \left|\mathbf{I}_{p} - \gamma \boldsymbol{\Sigma}\mathbf{B}\right|^{-1} .
    \end{equation}
  \end{Lemma}

  Define the event $\mathcal{G}_\delta =\bigcup_{S_1} \left\{ \sum_{i=1}^{n}  {\chi_i^2(2L)}  \leq  2nL(1+\delta) \right\}$ for $\delta\geq0$.
  Similar to~\eqref{eq:noCSI_pft2_goodregion}, we can bound $\mathbb{P}  \left[ \mathcal{G}_{\omega,\nu}^c \right]$ as
  \begin{equation} \label{eq:noCSI_q2t_goodregion}
    \mathbb{P}  \left[ \mathcal{G}_{\omega,\nu}^c \right]
    \leq \min_{\delta\geq 0}
    \left\{  \mathbb{P}  \left[ \mathcal{G}_{\omega,\nu}^c \cap \mathcal{G}_\delta \right]
    + \mathbb{P} \left[ \mathcal{G}_\delta^c \right]  \right\}.
  \end{equation}
  Here, $\mathbb{P} \left[ \mathcal{G}_\delta^c \right] \!=\! 1 \!-\! \frac{\gamma\left( nL, nL \left( 1+\delta \right)\right)}{\Gamma\left( nL \right)}$
  and $\mathbb{P} \! \left[ \mathcal{G}_{\omega,\nu}^c \!\cap\! \mathcal{G}_\delta \right]$ is bounded~as
    \begin{align}
      &\mathbb{P}  \left[ \mathcal{G}_{\omega,\nu}^c \cap \mathcal{G}_{\delta} \right] \notag \\
      & = \!\mathbb{P} \! \left[ \bigcup_{S_1} \left\{ {\sum}_{l=1}^{L} \! \left( {\mathbf{y}}_l^H \!\left( \mathbf{F}^{-1}  - \omega  \mathbf{F}_1^{-1} \right) {\mathbf{y}}_l \right) > b \right\} \cap \mathcal{G}_\delta \right]\label{eq:q2t_y} \\
      & = \!\mathbb{E}  \!\left[  \mathbb{P} \!\left[ \bigcup_{S_1} \left. \!\left\{ \sum_{i=1}^{n} \!\left(1\!-\!\omega\!-\!\omega \lambda_i\right) \!\frac{\chi_i^2(2L)}{2}
      \!>    b \right\}
      \cap \mathcal{G}_\delta \right| \!\mathbf{C}
      \right] \right] \! \label{eq:q2t_tildey}\\
      & \leq \! \sum_{S_1} \! \mathbb{E}\!\left[  \mathbb{P} \! \left[ \left.  \sum_{i=1}^{m} \! \frac{\lambda_i \chi_i^2(2L)}{2}  \!<\!  \frac{nL(1\!+\!\delta)(1\!-\!\omega) \!-\! b}{\omega}  \right| \!\mathbf{C}
      \right] \right] \label{eq:noCSI_q3t_cap} \\
      & \leq \binom{ K_a }{t} \mathbb{E} \! \left[
      \frac{\gamma\!\left( Lm, c_{\delta} \right)}{\Gamma\left( Lm \right)}  \right] , \label{eq:noCSI_q2t_prod}
    \end{align}
  where $b = L ( \omega \ln\!\left|\mathbf{F}_1\right| -  \ln\!\left|\mathbf{F}\right| + n \nu )$;
  \eqref{eq:q2t_tildey} holds because $\mathbf{y}_{l} \!=\!  \mathbf{F}^{\frac{1}{2}} \tilde{\mathbf{y}}_{l} \!\stackrel{\!\rm{i.i.d.}\!}{\sim}\! \mathcal{CN}\!\left( \mathbf{0}, \mathbf{F} \right)$ conditioned on $\mathbf{C}$ with $\tilde{\mathbf{y}}_{l} \!\!\stackrel{\!\rm{i.i.d.}\!}{\sim}\!\! \mathcal{CN}\!\left( \mathbf{0}, \mathbf{I}_n \right)$ for $l\in[L]$;
  \eqref{eq:noCSI_q3t_cap} follows from the union bound and the inequality $\mathbb{P}\left[\{Z \geq 0\} \cap \{W \geq 0\}\right] \leq \mathbb{P}\left[ u Z + r W \geq 0 \right]$ for $u,r\geq 0$;
  \eqref{eq:noCSI_q2t_prod} follows from~\cite[Eq.~(4.6a.17)]{quadratic_form1}.
  Substituting \eqref{eq:noCSI_q2t_prod} into \eqref{eq:noCSI_q2t_goodregion}, we derive an upper bound $q_{2,t}\!\left( \omega, \nu \right)$ on~$\mathbb{P} \! \left[ \mathcal{G}_{\omega,\nu}^c \right]$.
  Together with \eqref{eq:noCSI_pft2_goodregion} and \eqref{eq:noCSI_q1t_exp_y}, $p_t$ in \eqref{eq_noCSI_pt} is obtained.
  \end{IEEEproof}
  \end{Theorem}

  The scaling law established in~\cite{Caire1} provides some insights into the possible regimes, but fails to characterize the specific energy efficiency for a practical URA system.
  To this end, we derive a FBL achievability bound on the minimum required energy-per-bit in Theorem~\ref{Theorem_noCSI_achi}.
  A severe problem therein is that the error event is the union of many events.
  {To address this problem, we utilize Fano's ``good region'' technique in Lemma~\ref{Lemma:gr} to bound the probability $\mathbb{P} \left[ \mathcal{G}_{e} \right]$ in~\eqref{eq:noCSI_pft2_goodregion},
  where the ``good region'' $\mathcal{R}_{t,S_1}$ is selected as in~\eqref{eq:goodregion_noCSI}.
  For the received signal falling into this region, the cost function in~\eqref{eq:g_noCSI} corresponding to the truly transmitted codewords is small, and thus it is more likely to correctly decode rather than output a set of non-transmitted codewords based on the decoding principle in~\eqref{eq:decoderoutput_noCSI}.
  Moreover, our ``good region'' is parameterized by $\omega$ and $\nu$.
  By adjusting the two parameters, we can flexibly tune the shape of this region to find a tighter achievability bound.
  }

%

  In MIMO channels, the signals received over $L$ antennas should be jointly dealt with since they share the same sparse support.
  However, it is difficult to apply the projection decoder in~\cite{RAC_fading} to address this problem since the angle between the subspace spanned by $L$ received signals and the one spanned by $K_a$ codewords is involved to analyse.
  Alternatively, we leverage the ML decoder, where the cost function~\eqref{eq:g_noCSI} is easy to handle, at the price of requiring prior fading distribution.

  Apart from the achievability bound, we provide a converse bound on the minimum required energy-per-bit in Theorem~\ref{prop_converse_noCSI}.\begin{Theorem}\label{prop_converse_noCSI}
    The minimum required energy-per-bit for the URA model described in Section~\ref{section2} can be lower-bounded as
    \begin{equation} \label{eq:P_tot_conv_singleUE_EbN0}
      E^{*}_{b}(n,M,\epsilon) \geq \inf \frac{nP}{J}.
    \end{equation}
    Here, the $\inf$ is taken over all $P > 0$ satisfying that
      \begin{equation} \label{eq:P_tot_conv_singleUE1}
        J - \log_2 {K_a} \leq
        - \log_2 { \mathbb{P}\left[ \chi^2(2L) \geq (1+(n+1)P)r
        \right] } ,
      \end{equation}
      where $r$ is the solution of
      \begin{equation}\label{eq:P_tot_conv_singleUE2}
        \mathbb{P} \left[ \chi^2(2L) \leq r \right] = \epsilon .
      \end{equation}
    \begin{IEEEproof}
  In~\cite[Theorem~3]{noCSI_conv}, a converse bound was provided for the single-user case in MIMO channels with unknown CSI.
  In this part, we assume both the transmitted codewords of $K_a-1$ active users and their CSI are known to derive~the converse bound, which can be obtained based on~\cite[Theorem~3]{noCSI_conv} with the following changes:
  1)~we choose the auxiliary distribution $\prod_{l=1}^{L} \mathcal{CN}(0,\mathbf{I}_{n+1})$ for simplicity; 
  2)~we change $J$ in~\cite[Theorem~3]{noCSI_conv} to $J - \log_2 K_a$ and the reasons are as follows.
  Since all users share a common codebook, there may exist message collisions; thus, the number $B$ of different messages among the known transmitted messages of $K_a-1$ active users satisfies $1 \leq B \leq K_a-1$.
  Therefore, the decoder aims to output another $K_a - B$ possible messages to recover the message transmitted by the remaining active user.
  To derive a converse bound, we loosen the list size from $K_a - B$ to $K_a$ and the PUPE becomes $\mathbb{P} [ W_1 \notin \hat{\mathcal{W}} ]$.
  Using the meta-converse variation for list decoding~\cite[Theorem~4.1]{RAC_fading}, we modify the result in~\cite[Theorem~3]{noCSI_conv} by changing $J$ to $ \log_2 M/K_a$. 
  \end{IEEEproof}
\end{Theorem}

  Theorem~\ref{prop_converse_noCSI} holds for all codes but is derived based on the knowledge of the transmitted messages of $K_a-1$ active users and their CSI.
  It can be loose when $K_a$ is large because multi-user interference (MUI) is a significant bottleneck and CSI is difficult to obtain in this case.
  In contrast, Theorem~\ref{prop_converse_noCSI_Gaussian} considers the case with unknown CSI and unknown transmitted messages.
  Since it is tricky to analyse, we make a stronger assumption of Gaussian codebook,
  which reduces the converse result to a weaker ensemble converse and raises an open question about whether a tight converse bound can be derived under more general assumptions on the codebook.

\begin{Theorem}\label{prop_converse_noCSI_Gaussian}
    For the URA model described in Section~\ref{section2}, assuming that the codebook has i.i.d. Gaussian entries with $M>2K_a$ and $\epsilon\leq 1-{K_a}/{M}$,
    the minimum required energy-per-bit can be lower-bounded as
    \begin{equation} \label{eq:P_tot_conv_EbN0_Gaussian}
      E^{*}_{b}(n,M,\epsilon) \geq \inf \frac{nP}{J}.
    \end{equation}
    Here, the $\inf$ is taken over all $P > 0$ satisfying that
    \begin{align}
      & nL  \log_2 ( 1 + K_aP )
      - L p_{no} \mathbb{E} \!\left[ \log_2  \left| \mathbf{I}_{n} + {\mathbf{X}}_{ K_a} {\mathbf{X}}_{ K_a}^{ H}  \right|  \right]
      \notag\\
      & \geq  \left(1 - \epsilon\right) K_a \left( J - \log_2{K_a} \right) - K_a h_2 \left( \epsilon \right) ,
      \label{P_tot_conv_noCSI}
    \end{align}
    where $p_{no} = 1-{\binom{K_a}{2}}/{M}$ and $ {\mathbf{X}}_{K_a} \in \mathbb{C}^{n\times K_{a}}$ has i.i.d. $\mathcal{CN}(0,P)$ entries.

  \begin{IEEEproof}
  We assume w.l.o.g. that the active user set is $[K_a]$.
  Let $\mathbf{X}\in\mathbb{C}^{n\times M}$ be a codebook matrix.
  Denote $\bar{\mathbf{X}}_{\!K_aM} \!=\! \left[\mathbf{X},\ldots,\mathbf{X}\right] \!\in\! \mathbb{C}^{n\times K_aM}$~and~$\bar{\mathbf{X}}  \!=\! \operatorname{diag}\! \left\{\bar{\mathbf{X}}_{\!K_aM}, \ldots, \bar{\mathbf{X}}_{\!K_aM} \right\} \!\in\! \mathbb{C}^{nL\times K_aML}$.
  Let $\bar{\mathbf{H}}_{l}$ be a $K_a M  \times K_a M$ block diagonal matrix, whose block $k$ is a diagonal $M \!\times\! M$ matrix with diagonal entries equal to ${h}_{k,l}$.
  Let $\bar{\mathbf{H}}  \!=\! \left[\bar{\mathbf{H}}_{1}, \ldots , \bar{\mathbf{H}}_{L} \right]^T$.
  The vector $\bar{\boldsymbol{\beta}}  \!\in\! \left\{ 0,1 \right\}^{\!K_aM}$ has $K_a$ blocks, whose block $k$ denoted as $\bar{\boldsymbol{\beta}}_k$ is of size $M$ and includes one $1$. Then, we~have
  \begin{equation} \label{eq:P_tot_conv_y}
    \bar{\mathbf{y}} = \bar{\mathbf{X}}  \bar{\mathbf{H}}  \bar{\boldsymbol{\beta}} + \bar{\mathbf{z}} \in \mathbb{C}^{nL\times 1} ,
  \end{equation}
  where $\bar{\mathbf{z}} \in \mathbb{C}^{nL\times 1} $ with each entry i.i.d. from $\mathcal{CN}(0,1)$.

  Let $P_{e,k}^{de} = \mathbb{P} [ W_k \!\notin\! \hat{\mathcal{W}} ]$
  and $P_{e}^{de}\!=\!\frac{1}{K_a}\sum_{k\in [K_a]} \! P_{e,k}^{de} $.
  Since $P_{e}^{de}\!\leq \!P_e$, a converse bound based on the constraint $P_{e}^{de}\!\leq\!\epsilon$ is also converse for the constraint~\eqref{PUPE}.
  Applying~\cite[Eq.~(61) and Eq.~(63)]{RAC_fading} (where Eq.~(61) follows from Fano's inequality~\cite{elements_IT}) and allowing $S_2$ therein to be $[K_a]$ with
  $M>2K_a$, we have
  \begin{equation}
    \log_2\!\frac{M}{K_a}
    - h_2\!\left( P_{e}^{de} \right)
    - P_{e}^{de} \log_2\!\left( \!\frac{M}{K_a}\!-\!1 \!\right)
    \!\leq\! \frac{ I \big( \mathcal{W}_{[K_a]};\hat{\mathcal{W}} \left| \bar{\mathbf{X}} \right. \big) }{K_a}.\label{eq:fano_S}
  \end{equation}
  Assuming $P_{e}^{de} \leq \epsilon \leq 1- {K_a}/{M}$, we have
  \begin{equation}
    P_{e}^{de} \log_2\!\left( \!\frac{M}{K_a}-1 \right) + h_2 \left( P_{e}^{de} \right)
    \leq \epsilon \log_2\! \frac{M}{K_a} + h_2\left( \epsilon \right). \label{eq:fano_left2}
  \end{equation}

  The mutual information in~\eqref{eq:fano_S} is bounded as~\cite[Eq.~(149)]{finite_payloads_fading}
    \begin{equation} \label{eq_conv_fano_nocsi}
      I \;\big( \mathcal{W}_{[K_a]};\hat{\mathcal{W}} \left| \bar{\mathbf{X}} \right. \big)  \leq
      I \!\left(\left. \bar{\mathbf{H}}\bar{\boldsymbol{\beta}}; \bar{\mathbf{y}}  \right| \bar{\mathbf{X}} \right)
      - I \!\left( \left.\bar{\mathbf{H}}\bar{\boldsymbol{\beta}}; \bar{\mathbf{y}}  \right| \bar{\boldsymbol{\beta}}, \bar{\mathbf{X}}\right).
    \end{equation}
  Then, we can obtain
    \begin{align}
      I \left( \left. \bar{\mathbf{H}}\bar{\boldsymbol{\beta}}; \bar{\mathbf{y} } \right| \bar{\mathbf{X}} \right)
      & \leq L \mathbb{E} \left[  \log_2 \left| \mathbf{I}_{n} + \frac{1}{M} \bar{\mathbf{X}}_{K_aM} \bar{\mathbf{X}}_{K_aM}^{H}  \right|  \right]  \label{eq_conv_inf_term1_y}   \\
      & \leq nL \log_2 \left( 1 + K_aP \right) ,  \label{eq_conv_inf_term1}
    \end{align}
  where \eqref{eq_conv_inf_term1_y} follows from the inequality in~\cite[Eq.~(151)]{finite_payloads_fading} and
  \eqref{eq_conv_inf_term1} follows from Jensen's inequality assuming that the codebook has i.i.d. entries with mean $0$ and variance $P$.
  The term $I \left( \left.\bar{\mathbf{H}}\bar{\boldsymbol{\beta}}; \bar{\mathbf{y}}  \right| \bar{\boldsymbol{\beta}}, \bar{\mathbf{X}}\right)$ can be lower-bounded as follows:
  \vspace{-0.1cm}
    \begin{align}
      \!I\! \left( \left.\!\bar{\mathbf{H}}\bar{\boldsymbol{\beta}};\bar{\mathbf{y}} \right| \!\bar{\boldsymbol{\beta}},  \bar{\mathbf{X}}\right)
      & \!\geq\! L \!\;\!\mathbb{E} \!\left[ \log_2 \!\left| \mathbf{I}_{n} \!+\! \mathbf{X}
      {\sum}_{k\!=\!1}^{K_a} \!\! \operatorname{diag}(\bar{\boldsymbol{\beta}}_k)
      {\mathbf{X}}^{H} \right| \right] \!\label{eq_conv_inf_term2_yury} \\
      & \!\geq p_{no} \!\;L\!\; \mathbb{E} \left[ \log_2  \left| \mathbf{I}_{n} + \mathbf{X}_{K_a}  \mathbf{X}_{K_a}^{ H} \right|  \right] \label{eq_conv_inf_term2},
    \end{align}
    where $p_{no}$ denotes the probability of no collision 
    and $\mathbf{X}_{K_a}$ includes codewords transmitted by $K_a$ active users without collision.
    Here, 
    \eqref{eq_conv_inf_term2} holds because
    $\mathbb{E} \left[ f\left( \mathbf{X} , \bar{\boldsymbol{\beta}} \right) \right]
    = p_{no} \mathbb{E} \left[ f_{\text{no-col}}\left( \mathbf{X} , \bar{\boldsymbol{\beta}} \right) \right]
    + (1-p_{no}) \mathbb{E} \left[ f_{\text{col}}\left( \mathbf{X} , \bar{\boldsymbol{\beta}} \right) \right]$,
    where the expectation of $f_{\text{no-col}}\left( \mathbf{X} , \bar{\boldsymbol{\beta}} \right) = \log_2  \left| \mathbf{I}_{n} + \mathbf{X}_{K_a}  \mathbf{X}_{K_a}^{ H} \right|$ can be evaluated under the assumption of Gaussian codebook.
    It completes the proof of Theorem~\ref{prop_converse_noCSI_Gaussian}.
  \end{IEEEproof}
\end{Theorem}


\section{Numerical Results} \label{sec_simulation}
\begin{figure}
    \centering
    \includegraphics[width=0.88\linewidth]{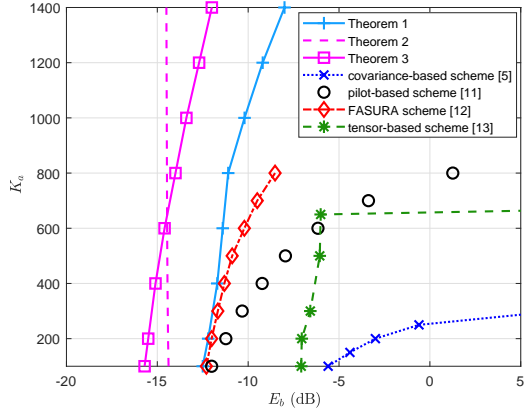}\\
    \vspace{-0.1cm}
    \caption{The number of active users versus the energy-per-bit for URA with $n=3200$, $J = 100$~bits, $L=50$, $\epsilon = 0.1$ for the tensor-based scheme, and $\epsilon = 0.025$ otherwise.}
  \label{fig:L32}
  \end{figure}

  In this section, we provide numerical evaluation of the derived bounds.
  In Fig.~\ref{fig:L32}, we consider the scenario with $n=3200$, $J=100$ bits, $L = 50$, and $\epsilon\in\{0.025,0.1\}$.
  In this case, we compare our achievability and converse bounds, as well as the schemes proposed in~\cite{Caire1,Caire2,Fasura,TBM},
  in terms of the minimum required energy-per-bit for different numbers of active users.
  Next, we explain how each curve is obtained:
  \begin{enumerate}
    \item For the achievability bound is Theorem~\ref{Theorem_noCSI_achi}, we generate $10000$ samples to evaluate the expectations therein using the Monte Carlo method.

    \item The converse bounds in Theorem~\ref{prop_converse_noCSI} and Theorem~\ref{prop_converse_noCSI_Gaussian} are plotted, respectively.
       The expectations therein are evaluated by the Monte Carlo method using $500$ samples.

    \item To evaluate the covariance-based scheme proposed in~\cite{Caire1}, we adopt a frame of $16$ slots each with $200$ dimensions.
        The binary subblock length is $15$ with the parity profile $[0,7,8,8,9, \ldots, 9, 13 , 14]$ for the tree code.
        We obtain the average of the probabilities of~misdetection and false-alarm, i.e. $P_e = \left(p^{md} +  p^{fa}\right)/2$, and plot the minimum required energy-per-bit to satisfy $P_e \leq \epsilon$.

    \item The pilot-based scheme is evaluated as in~\cite[Fig.~7]{Caire2}. The data is split into two parts with $16$~bits and $84$~bits, respectively. The first part is coded as ``pilot'' of length $1152$; the second one is coded by a polar code of length $2048$.
        The error requirement is the same as that for the covariance-based scheme.

    \item The FASURA scheme is evaluated as in~\cite[Fig.~4]{Fasura}. Similar to the pilot-based scheme in~\cite{Caire2}, it divides~messages into two parts.
        Departures from~\cite{Caire2} include the use of spreading sequences, the detection of active sequences, and channel/symbol estimation techniques.


    \item The tensor-based scheme proposed in~\cite{TBM} is evaluated with tensor signature $(8,5,5,4,4)$, an outer BCH code, and a higher error requirement $\epsilon = 0.1$.
  \end{enumerate}
  As predicted before, we can observe from Fig.~\ref{fig:L32} that the converse bound in Theorem~2 dominates in small $K_a$ regime~and the weaker ensemble bound in Theorem~3 dominates otherwise.
  Numerical results verify the tightness of our bounds, with the gap between the achievability and (dominating) converse bounds less than $5$~dB in our regime.
  In MIMO channels, the required $E_b$ is almost a constant when $K_a$ is small, in line with the almost perfect MUI cancellation in the single-receive-antenna setting~\cite{A_perspective_on,RAC_fading}.
  Moreover, our bounds provide theoretical benchmarks to evaluate practical schemes. Specifically, among the schemes proposed in~\cite{Caire1,Caire2,TBM,Fasura}, the FASURA scheme in~\cite{Fasura} performs the best, but it still exhibits a large gap to our bounds in large $K_a$ regime.
  How to reduce this gap is an interesting topic for the future work.

\begin{figure}
    \centering
    \includegraphics[width=0.88\linewidth]{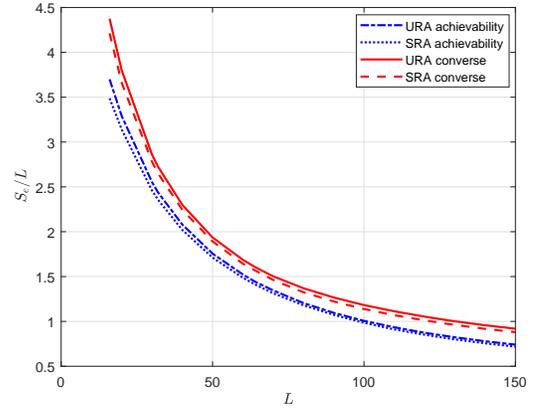}\\
    \vspace{-0.1cm}
    \caption{The spectral efficiency per antenna versus the number of BS antennas for URA and SRA ($K_a/K=0.4$) with $n=1000$, $J = 100$~bits, $E_b=16$~dB, and $\epsilon = 0.1$.}
  \label{fig:SE_noCSI}
  \end{figure}

  In Fig.~\ref{fig:SE_noCSI}, we compare the achievability bound (in Theorem~\ref{Theorem_noCSI_achi}) and the converse bound (in Theorem~\ref{prop_converse_noCSI_Gaussian}) for URA in terms of the maximum spectral efficiency per antenna for different numbers of BS antennas with $n=1000$, $J = 100$~bits, $E_b=16$~dB, and $\epsilon = 0.1$.
  Since both $n$ and $J$ are fixed, Fig.~\ref{fig:SE_noCSI} indicates the bounds on the number of reliably served active users against that of BS antennas.
  We also show the theoretical bounds for SRA provided in~\cite{TIT}.
  Compared with SRA, the URA paradigm achieves higher spectral efficiency
  because all users share a common codebook and the search space to find the transmitted messages is reduced in this case.
  For both SRA and URA paradigms, the total spectral efficiency $S_e$ increases with $L$, whereas the spectral efficiency per antenna $S_e/L$ gradually reduces due to the increased channel uncertainty.

\section{Conclusion} \label{section:conclusion}
  In this paper, we considered the massive URA problem in MIMO quasi-static Rayleigh fading channels with stringent latency and energy constraints.
  Specifically, an achievability bound, a general converse bound, and a weaker ensemble converse bound were derived under a PUPE constraint.
  Numerical evaluation verified the tightness of our results and indicated that some existing schemes exhibit a large gap to our theoretical bounds, especially when there are many users.
  Moreover, we observed that compared with the SRA paradigm, the URA paradigm achieves higher spectral efficiency. 


\begin{thebibliography}{00}


\bibitem{wuyp} Y. Wu, X. Gao, S. Zhou, W. Yang, Y. Polyanskiy, and G. Caire, ``Massive access for future wireless communication systems,'' \emph{IEEE Wireless Commun.}, vol. 27, no. 4, pp. 148--156, Aug. 2010.

\bibitem{A_perspective_on} Y. Polyanskiy, ``A perspective on massive random-access,'' in \emph{Proc. IEEE Int. Symp. Inf. Theory (ISIT)}, Aachen, Germany, Jun. 2017, pp. 2523--2527.

\bibitem{RAC_fading} S. S. Kowshik, K. Andreev, A. Frolov, and Y. Polyanskiy, ``Energy efficient coded random access for the wireless uplink,'' \emph{IEEE Trans. Commun.}, vol. 68, no. 8, pp. 4694--4708, Aug. 2020.

\bibitem{noKa} K.-H. Ngo, A. Lancho, G. Durisi, and A. Graell i Amat, ``Unsourced multiple access with random user activity,'' Feb. 2022, arxiv:2202.06365. [Online]. Available: https://arxiv.org/abs/2202.06365



\bibitem{Caire1} A. Fengler, S. Haghighatshoar, P. Jung, and G. Caire, ``Non-bayesian activity detection, large-scale fading coefficient estimation, and unsourced random access with a massive MIMO receiver,'' \emph{IEEE Trans. Inf. Theory}, vol. 67, no. 5, pp.~2925--2951, May 2021.

\bibitem{1961} R. M. Fano, \emph{Transmission of Information}. Jointly published by the MIT Press and John Wiley $\&$ Sons, 1961.

\bibitem{elements_IT} T. M. Cover and J. A. Thomas, \emph{Elements of Information Theory}, 2nd ed. Hoboken, NJ, USA: John Wiley $\&$ Sons, 2006.

\bibitem{quadratic_form1} A. M. Mathai and B. P. Serge, \emph{Quadratic Forms in Random Variables: Theory and Applications}. New York, NY, USA: Marcel Dekker, 1992.


\bibitem{noCSI_conv} J. {\"O}stman, W. Yang, G. Durisi, and T. Koch, ``Diversity versus multiplexing at finite blocklength,'' in \emph{Proc. IEEE Int. Symp. Wireless Commun. Syst. (ISWCS)}, Barcelona, Spain, Aug. 2014, pp. 702--706.






\bibitem{finite_payloads_fading} S. S. Kowshik and Y. Polyanskiy, ``Fundamental limits of many-user MAC with finite payloads and fading,'' \emph{IEEE Trans. Inf. Theory}, vol. 67, no. 9, pp. 5853--5884, Sep. 2021.
\bibitem{Caire2} A. Fengler, O. Musa, P. Jung, and G. Caire, ``Pilot-based unsourced random access with a massive MIMO receiver, interference cancellation, and power control,'' \emph{IEEE J. Sel. Areas Commun.}, vol. 40, no. 5, pp. 1522--1534, May 2022.
\bibitem{Fasura} M. Gkagkos, K. R. Narayanan, J. -F. Chamberland, and C. N. Georghiades, ``FASURA: A scheme for quasi-static massive MIMO unsourced random access channels,'' in \emph{Proc. IEEE Int. Workshop Signal Process. Adv. Wireless Commun. (SPAWC)}, Oulu, Finland, Jul. 2022.
\bibitem{TBM} A. Decurninge, I. Land, and M. Guillaud, ``Tensor-based modulation for unsourced massive random access,'' \emph{IEEE Wireless Commun. Lett.}, vol. 10, no. 3, pp. 552--556, Mar. 2021.

\bibitem{TIT} J. Gao, Y. Wu, S. Shao, W. Yang, and H. V. Poor, ``Energy efficiency of massive random access in MIMO quasi-static Rayleigh fading channels with finite blocklength,'' to appear in \emph{IEEE Trans. Inf. Theory}, Nov. 2022, arxiv:2210.11970. [Online]. Available: https://arxiv.org/abs/2210.11970




\end{thebibliography}
\end{document}